\documentclass[%
 reprint,
superscriptaddress,
 amsmath,amssymb,
 aps,
]{revtex4-1}

\usepackage{graphicx}
\usepackage{dcolumn}
\usepackage{bm}
\usepackage{ulem}
\usepackage{hyperref}

\begin{document}


\title{First-principles studies of the local structure and relaxor behavior of Pb(Mg$_{1/3}$Nb$_{2/3}$)O$_3$$-$PbTiO$_3$-derived ferroelectric perovskite solid solutions}

\author{Hengxin Tan}
\affiliation{State Key Laboratory of Low-Dimensional Quantum Physics and Collaborative Innovation Center of Quantum Matter, Department of Physics, Tsinghua University, Beijing 100084, China}
\affiliation{Department of Chemistry, University of Pennsylvania, Philadelphia, Pennsylvania 19104-6323, USA}

\author{Hiroyuki Takenaka}
\affiliation{Department of Chemistry, University of Pennsylvania, Philadelphia, Pennsylvania 19104-6323, USA}

\author{Changsong Xu}
\affiliation{State Key Laboratory of Low-Dimensional Quantum Physics and Collaborative Innovation Center of Quantum Matter, Department of Physics, Tsinghua University, Beijing 100084, China}

\author{Wenhui Duan}
\affiliation{State Key Laboratory of Low-Dimensional Quantum Physics and Collaborative Innovation Center of Quantum Matter, Department of Physics, Tsinghua University, Beijing 100084, China}
\affiliation{Institute for Advanced Study, Tsinghua University, Beijing 100084, China}

\author{Ilya Grinberg}
\affiliation{Department of Chemistry, Bar-Ilan University, Ramat-Gan, 52900, Israel}

\author{Andrew M. Rappe}
\email{rappe@sas.upenn.edu}
\affiliation{Department of Chemistry, University of Pennsylvania, Philadelphia, Pennsylvania 19104-6323, USA}


\begin{abstract}
  We have investigated the effect of transition metal dopants on the local structure of the prototypical 0.75Pb(Mg$_{1/3}$Nb$_{2/3}$)O$_3$$-$0.25PbTiO$_3$ relaxor ferroelectric.
  We find that these dopants give rise to very different local structure and other physical properties.
  For example, when Mg is partially substituted by Cu or Zn, the displacement of Cu or Zn is much larger than that of Mg, and is even comparable to that of Nb.
  The polarization of these systems is also increased, especially for the Cu-doped solution, due to the large polarizability of Cu and Zn.
  As a result, the predicted maximum dielectric constant temperatures ($T_{\textrm{m}}$) are increased.
  On the other hand, the replacement of a Ti atom with a Mo or Tc dramatically decreases the displacements of the cations and the polarization, and thus, the $T_{\textrm{m}}$ values are also substantially decreased.
  The higher $T_{\textrm{m}}$ cannot be explained by the conventional argument based on the ionic radii of the cations.
  Furthermore, we find that Cu, Mo, or Tc doping increase the cations displacement disorder.
  The effect of the dopants on the temperature dispersion $\Delta$$T_{\textrm{m}}$, which is the change of $T_{\textrm{m}}$ for different frequencies, is also discussed. Our findings lay the foundation for further investigations of unexplored dopants.

\begin{description}
\item[PACS numbers]
77.80.-e, 71.15.Mb.
\end{description}
\end{abstract}



\maketitle

\section{\label{sec:level1}Introduction}

In the past decades, perovskite (ABO$_3$) relaxor ferroelectrics \cite{JAP111p031301, JAD2p1241010, PMS65p124} have been extensively studied, due to their use in capacitors and piezoelectric devices such as SONAR and medical ultrasound imaging.
Unlike normal ferroelectrics which exhibit a very narrow peak vs. temperature and no frequency dependence for their dielectric response, relaxor ferroelectrics show a broad and frequency-dependent response around it's temperature maximum $T_{\textrm{m}}$.
Typical relaxor ferroelectrics such as Pb(Mg$_{1/3}$Nb$_{2/3}$)O$_3$ (PMN) and Pb(Zn$_{1/3}$Nb$_{2/3}$)O$_3$ (PZN) \cite{Shirane-arXiv} are technologically important due to their good electromechanical properties.
Moreover, it was found that their solid solutions with PbTiO$_3$ (PT), i.e. PMN-PT and PZN-PT, exhibit extremely large electromechanical coupling factor and piezoelectric coefficient $d_{33}$ \cite{JAP82p1804,JAP85p2810,JAP90p3471} around the morphotropic phase boundary (MPB) \cite{MPB}.
This has led to the intense interest in these materials.
While many experimental investigations have focused on the dielectric dispersion and piezoelectric properties \cite{PRB66p094112,JACS86p1861,APL87p012904,nature441p956,JPD43p015405,JAP113p187208}, the connection of composition and relaxor behavior is still not fully understood, even though it is known that heterovalency or a degree of disorder on the B-site is essential.
However, by capturing the key features of the local atomic environment, our previous studies \cite{Ilya04p144118, Ilya04p220101, Ilya05p094111, Ilya07p267603, Qi10p134113, Hiro14p1, Hiro13p147602, Hiro17p391} using both density functional theory (DFT) and molecular dynamics (MD) have shed light on the relationship between the local structure and the relaxor behavior, as well as on the dynamics of relaxors.
Grinberg $et$ $al$. had also shown the strong dipole-dipole scatter in PbZr$_x$Ti$_{1-x}$O$_3$ (PZT) \cite{Ilya04p144118} and PMN \cite{Ilya04p220101} by DFT calculations in 2004.
Thus, properties such as the MPB location and the temperature dispersion (e.g. $\Delta$$T_{\textrm{m}}$ = $T_{\textrm{m}}$(1 MHz)-$T_{\textrm{m}}$(100 Hz)) that measures the strength of relaxor behavior can be predicted from composition \cite{Ilya07p267603}.

In addition to binary compounds such as PMN-PT and PZN-PT, current research interest has been directed to doped PMN-PT materials such as Mn-, Fe-, Zn- and W-doped relaxor solid solutions \cite{APL89p162906, APL84p4711, JAP113p204101, JAP101p114107}, because such doped systems may exhibit enhanced properties relative to the undoped solid solutions. Ternary compounds such as Pb(In$_{1/2}$Nb$_{1/2}$)O$_3$$-$Pb(Mg$_{1/3}$Nb$_{2/3}$)O$_3$$-$PbTiO$_3$ (PIN-PMN-PT) \cite{PRB90p134107, APL107p082902} and Pb(Zn$_{1/3}$Nb$_{2/3}$)O$_3$$-$Pb(Mn$_{1/3}$Nb$_{2/3}$)O$_3$$-$PbTiO$_3$ (PZN-PMN-PT) \cite{JMR10p3185, CI38p3835} have also attracted intense interest. Such compounds are much more complex than binary compounds and can exhibit different phases depending on the different end member compositions. A typical phase diagram for Pb(Mg$_{1/3}$Nb$_{2/3}$)O$_3$$-$PbZrO$_3$$-$PbTiO$_3$ (PMN-PZ-PT or PMN-PZT) is given in Ref.\cite{JACS48p630}, where the curves of the MPB are also shown.

Nevertheless, to the best of our knowledge, almost all investigations of either doped or ternary systems have been experimental, and there have been almost no DFT investigations focusing on the relationship between the local structure and the properties of such compounds. Therefore, in this study, we investigate the microscopic structures using the first-principles method for several systems created by substituting B-site atoms in a PMN-PT composition with transition metals.
The substitutions used in this work do not change the tolerance factors $t$ \cite{Goldschmidt} (Table I) very much, and all systems are still stable after doping. The widely-used tolerance factor $t$ for the perovskite structure is defined as \[t=\frac{R_{\textrm{A}}+R_{\textrm{O}}}{\sqrt{2}\left(R_{\textrm{B}}+R_{\textrm{O}}\right)} ,\]
where \(R_{\textrm{A}}\), \(R_{\textrm{B}}\) and \(R_{\textrm{O}}\) are the ionic radii \cite{AC32p751} of A-site, B-site and oxygen ions, respectively. Generally, a perovskite with a tolerance factor lower than 0.8 or higher than 1.1 is not stable due to the mismatch between the preferred A-O and B-O sublattice sizes.
We find that transition metal dopants affect the local Pb displacements which contribute substantially to the total polarization of Pb-based perovskites.
This also means that the temperature-dependent physical properties in Pb-based relaxors are sensitive to the changes induced by B-site dopants.
We therefore estimate the effect of the doping on the temperature $T_{\textrm{m}}$ of the dielectric maximum in this work.
Our results show that Cu (ionic radius of 0.73\AA) substitution for Mg (0.72\AA) significantly increases $T_{\textrm{m}}$, compared to other systems explored in this work, and indicate that the origin of the high $T_{\textrm{m}}$  does not arise from its ionic radius.

Using first-principles DFT calculations, we report the substitution effects on the local environment characteristics including cation displacements, polarization, and relaxor behavior. The paper is organized as follows. In Sec. II, we provide the details of the structural model used in our calculations, and describe the computational methods. Section III presents the results of our calculations, including the local structure, electronic structure and the maximum dielectric constant temperatures $T_{\textrm{m}}$. We finally summarize this work in Sec. IV.

\section{\label{sec:leve1}MODEL AND METHODOLOGY}

Among the PMN-based solid solutions, those with PMN content near 75\% and PT content near 25\% (0.75PMN-0.25PT) have been intensively studied \cite{Hiro13p147602, Hiro17p391, APL77p1888, JAP97p094107, JAP100p124112, JPCM12pL541, JPCM23p435902}.
Davies $et$ $al$. \cite{JPCS61p159} showed that the B-cation arrangement in PMN follows the random-site model, with B$^{'}$ and B$^{''}$ sites in Pb(B$^{'}_{1/2}$B$^{''}_{1/2}$)O$_3$ in the rock-salt arrangement.
In such a model, 0.75PMN-0.25PT has full Nb occupation of the B$^{'}$ site, and equal Mg and Ti occupation of the B$^{''}$ site (Fig.1).
This is a very convenient choice for first-principles studies, because it allows the use of a relatively small supercell, and we therefore take this composition and B-cation ordering as the basic system in our studies.
A 2$\times$2$\times$2 40-atom supercell of 0.75PMN-0.25PT (abbreviated as PMN-PT hereafter) is employed throughout this work.
This supercell is large enough to capture the local structural characteristics, as shown in previous studies \cite{Ilya04p144118,Ilya05p094111}.
Seven doped systems are investigated, with four systems created by replacing half of the Mg with Fe, Ni, Cu or Zn, and the other three systems created by replacing half of the Ti atoms with Zr, Mo or Tc.
These doped systems are thus abbreviated hereafter as PMN-PT-Fe/Ni/Cu/Zn/Zr/Mo/Tc, respectively.

DFT calculations are performed using the Quantum ESPRESSO software package \cite{JPCM21p395502}.
The structures are fully optimized using the quasi-Newton algorithm with no symmetry imposed.
All Hellman-Feynman forces are converged to less than 10$^{-4}$ Ry/Bohr and the total energies are converged to less than 10$^{-6}$ Ry/cell.
The designed non-local, optimized norm-conserving pseudopotentials \cite{Rappe90p1227, Ramer99p12471} are generated with the OPIUM code \cite{web}.
The local density approximation (LDA) exchange-correlation functional and a 4$\times$4$\times$4 Monkhorst-Pack $k$-mesh \cite{PRB13p5188} are used in all calculations, except for the polarization calculations which are performed using the Berry-phase method and the denser 6$\times$6$\times$6 grid.
The plane-wave basis set cutoff is 60 Ry in all calculations.
The Hubbard $U$ correction is used for partially-filled $d$ orbitals, with $U$ of 4 eV for Fe \cite{PRB83P094105}, 7 eV for Ni \cite{PRB44p943}, 6.52 eV for Cu \cite{PRB44p943}, 2.59 eV for Mo \cite{PRB86p165105} and 2.49 eV for Tc \cite{PRB86p165105}, while no correction is used for Ti, Zn, Zr and Nb, for which the $d$ orbitals are either fully filled (for Zn$^{2+}$) or empty (for Ti$^{4+}$, Zr$^{4+}$ and Nb$^{5+}$).

\begin{figure}[htb!]
 \centering
 \includegraphics[width=0.95\columnwidth]{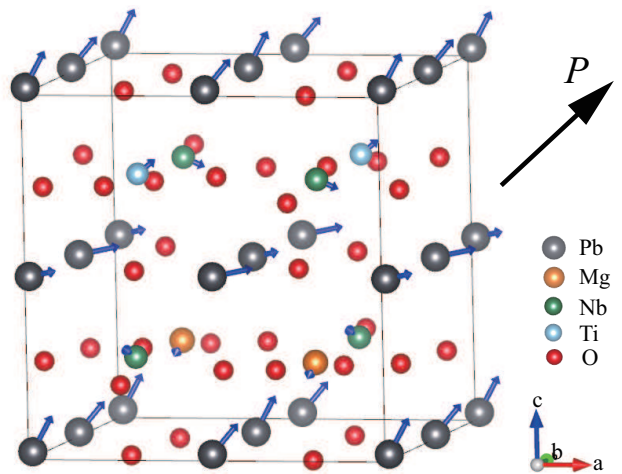}
 \caption{\label{Fig1}(Color online) Schematic of PMN-PT structure. The cation displacements away from the high-symmetry positions are indicated by blue arrows, scaled up by a factor of three for better visualization. The bold black arrow indicates the direction of the polarization. Note that the doped systems are created by replacing one of the two Mg atoms or one of the two Ti atoms in each supercell with a substituent atom.}
\end{figure}

\section{\label{sec:leve1}RESULTS AND DISCUSSIONS}
\subsection{\label{sec:level2}Local structure}

\begin{table*}
\caption{Local structure parameters of PMN-PT and doped systems. $t$ is the tolerance factor. $V$ is the volume of the 2$\times$2$\times$2 40-atom supercell in \AA$^3$. $D_\textrm{Pb}$, $D_\textrm{Mg}$, $D_\textrm{Nb}$, $D_\textrm{Ti}$ and $D_\textrm{X}$ are the average displacements of Pb, Mg, Nb, Ti and X cations respectively, where X stands for the dopant element, and $D_\textrm{FE,B}$ is the weighted average displacement of the FE-active B-site cations. All displacements are in \AA. $\theta$$_\textrm{Pb}$ is the average scatter angle between the displacement vectors of any two Pb cations, in degrees ($^\circ$). $P$ is the polarization in $\mu$C/cm$^2$, and $T_\textrm{m}$ (K) represents the temperature of the maximum dielectric constant calculated using equation (3) in Ref.\cite{Ilya04p220101} at 1 MHz \cite{note1}.}
\renewcommand\arraystretch{1.30}
\begin{ruledtabular}
\begin{tabular}{lccccccccccccc}
  Systems & $t$ & $V$ & $D_\textrm{Pb}$ & $D_\textrm{Mg}$ & $D_\textrm{Nb}$ & $D_\textrm{Ti}$ & $D_\textrm{X}$ & $D_\textrm{FE,B}$ & $\theta$$_\textrm{Pb}$ & $P$ & $T_\textrm{m}$\\
\hline
  PMN-PT    & 0.9962 & 509.1 & 0.368 & 0.066 & 0.173 & 0.233 & -     & 0.193 & 35.5 & 62.4 & 463 \\
  PMN-PT-Fe & 0.9926 & 512.2 & 0.362 & 0.067 & 0.175 & 0.239 & 0.068 & 0.196 & 21.7 & 61.6 & 451 \\
  PMN-PT-Ni & 0.9981 & 507.5 & 0.352 & 0.062 & 0.166 & 0.226 & 0.045 & 0.186 & 26.7 & 59.5 & 421 \\
  PMN-PT-Cu & 0.9956 & 510.2 & 0.421 & 0.077 & 0.195 & 0.233 & 0.146 & 0.199 & 40.4 & 67.5 & 542 \\
  PMN-PT-Zn & 0.9950 & 508.1 & 0.377 & 0.065 & 0.175 & 0.230 & 0.151 & 0.187 & 28.7 & 64.5 & 495 \\
  PMN-PT-Zr & 0.9893 & 519.2 & 0.388 & 0.078 & 0.148 & 0.260 & 0.110 & 0.170 & 33.7 & 58.5 & 407 \\
  PMN-PT-Mo & 0.9935 & 510.9 & 0.316 & 0.063 & 0.128 & 0.216 & 0.074 & 0.146 & 43.3 & 49.9 & 296 \\
  PMN-PT-Tc & 0.9938 & 507.7 & 0.276 & 0.055 & 0.120 & 0.216 & 0.039 & 0.139 & 49.9 & 43.6 & 226 \\
\end{tabular}
\end{ruledtabular}
\end{table*}

The local structures of all systems are examined, and the results are summarized in Table I.
The results for PMN-PT are in good agreement with previous investigations \cite{Ilya04p220101, JPCM23p435902}.
For example, our calculations yield the polarization of 62.4 $\mu$C/cm$^2$ and the average Pb displacement scatter angle $\theta$$_{\textrm{Pb}}$ of 35.5$^\circ$, showing reasonable good agreement with the theoretical values of 55 $\mu$C/cm$^2$ and 33$^\circ$ in Ref.\cite{Ilya04p220101}.
The volume of PMN-PT is 509.1 \AA$^3$ per 40-atom supercell in our calculations.
This volume corresponds to 63.6 \AA$^3$ per formula unit, agreeing very well with the result of MD simulations using a shell model potential (about 63.5 \AA$^3$) by Sepliarsky $et$ $al$. \cite{JPCM23p435902}.
However, our calculated volume is about 2\% smaller than the experimental value (Ref.\cite{JPCM23p435902} and references therein), due to the usual LDA underestimate of the volume by 1-3\%.

We now focus on the local structure of PMN-PT.
Structural optimizations reveal that not only the average displacements of Pb and B-site atoms from the negative charge centers of the oxygen cages vary among the different systems, but also that doping can change the Pb displacement directions.
Inspection of Fig.1 shows that Mg atoms have a strong effect on the Pb displacement direction.
There are three types of B-site atom cube faces, namely the  Mg-Nb cube face ((001) plane), the Ti-Nb cube face ((001) plane) and the Mg-Nb-Ti cube face ((100) and (010) planes).
Pb tends to displace toward an Mg-Nb face and avoid a Ti-Nb face.
This stems from the bond order difference of the cation-oxygen bonds in such faces.
The oxygen atoms between Ti and Nb have higher B-O bond order compared to the oxygen atom between Mg and Nb.
The overbonded oxygen atoms create a strong repulsive interaction between Pb and the Ti-Nb face and thus impede the distortion of Pb toward the Ti-Nb face.
As a result, Pb displacements have small $c$ components toward the Ti-Nb face but relatively large $c$ components toward the Mg-Nb face, while the average Pb displacement direction is approximately [323].

Doped PMN-PT systems can show quite different local structure effects.
The doping of Fe and Ni into PMN-PT affects all displacements very weakly.
In these systems, one of the two Mg atoms in the supercell is substituted by one Fe or Ni.
The very small displacements of Fe and Ni are comparable to that of Mg.
Actually, we find that Fe and Ni play a similar role to Mg for the local structure in PMN-PT.
For example, the average displacement direction of Pb, which is similar to the overall [323] polarization direction, is hardly changed by Fe or Ni.
Moreover, bond lengths between the B-site atoms and oxygen atoms are also almost unchanged, as can be seen in Figs.S1 (b) and (c) of the Supplementary Material (SM), where all B-O bond lengths in all eight systems are shown.
At the same time, we also observe that the average scatter angle $\theta$$_\textrm{Pb}$ between Pb displacements in these two doped systems decreases slightly.
Such small changes in displacements, scatter angles and the cation-oxygen bond lengths reveal that these dopants have a modest effect on the local structure of PMN-PT.


In PMN-PT-Cu and PMN-PT-Zn, one of the two Mg atoms in the supercell is substituted with one Cu or Zn.
As shown in Table I, the displacements of Cu and Zn are much larger than the displacement of Mg and are even comparable to that of Nb.
Moreover, the variations of the Cu-O or Zn-O bond lengths are also much larger than those of the Mg-O bond lengths, as shown in Figs.S1 (d) and (e) of the SM.
The shorter and longer bond lengths are due to the polarizability of Cu and Zn.
As had been shown in Ref.\cite{Ilya04p220101}, Zn displacement arises from the covalent bonding of Zn with O due to the imperfect screening of the inner core charge by the localized $d$ electrons and the shallow $p$ orbitals \cite{Nature92p136}.
The shortened bonds also compensate the bond valence \cite{CR109p6858} of the underbonded oxygen.
The 3$d$ orbitals of Cu$^{2+}$ are partially occupied by 9 electrons that may not screen the inner core as well as the fully occupied 3$d$ orbitals of Zn$^{2+}$.
However, Cu$^{2+}$ has a greater polarizability and attracts adjacent Pb more strongly than Zn$^{2+}$.

The dopants on the Mg-site in PMN-PT give rise to significantly different local structure behaviors, even though the Fe$^{2+}$, Ni$^{2+}$, Cu$^{2+}$, and Zn$^{2+}$ ions have the ionic radii of 0.78~\AA, 0.69~\AA, 0.73~\AA, and 0.74~\AA, respectively, which are very close to the ionic radius of 0.72~\AA\ of Mg.
We expected that the Cu and Zn dopants would not show large displacements because the large radius of Mg relative to those of Nb and Ti in pure PMN-PT gives rise to essentially no off-centering.
It is observed that Cu has large displacements along the $a$-axis, intermediate displacements along the $c$-axis, and no displacement along the $b$-axis, indicating covalent bonding with nearest O atoms along the $x$ and $z$ directions.
Presumably, the large difference between Ni and Cu is due to the Jahn-Teller effect rather than the screening effect.
In a perfectly ordered system, distortions of octahedral cages along the $c$-axis lowers the energy, due to the strong Jahn-Teller effect induced by cations with nine 3$d$ electrons.
Since the cation and dipole disorder are strong in systems like PMN-PT, the octahedral cages are distorted causing the $d$ electrons to partially occupy the $d_{x^2-y^2}$ and $d_{z^2}$ states, and this gives rise to the covalent bonding.
By contrast, the eight 3$d$ electrons in Ni$^{2+}$ do not give rise to the Jahn-Teller distortions.
If this is true, the small difference between Fe and Ni in Figs.S1 (b) and (c) in SM arises from the weak Jahn-Teller effect due to six 3$d$ electrons in Fe$^{2+}$, which do not provide a sufficient driving force for the off-centering.
Furthermore, only the stronger distortions in the Cu-doped PMN-PT enhance the average scatter angle $\theta$$_\textrm{Pb}$ among the studied systems with doping on the Mg-site.

Unlike the Fe/Ni/Cu/Zn-doped systems discussed above, in PMN-PT-Zr/Mo/Tc, one of the two Ti atoms in the supercell is replaced.
Zr is not as ferroelectrically (FE)-active as Ti.
The displacement of Zr is much smaller than that of Ti (0.110 vs. 0.260~\AA), and the Zr-O bond lengths are more uniform compared to Ti-O bond lengths (See Fig.S1 (f) of the SM).
However, the Zr$^{4+}$ ionic radius (0.72~\AA) is much larger than that of Ti$^{4+}$ (0.605~\AA).
Thus, compared to PMN-PT, the volume of PMN-PT-Zr expands, resulting in greater space for cation displacements.
The competition between these different effects gives rise to slightly increased cation displacements, as shown in Table I, with the  exception of Nb, for which the displacement is slightly decreased due to the fact that the octahedron centered on Nb is compressed by the neighboring, larger ZrO$_6$ octahedron.
We also find that Zr has a very limited effect on the Pb displacement directions, as indicated by the almost unchanged $\theta$$_\textrm{Pb}$ value (33.7$^\circ$).
The situation is different for Mo- and Tc-doped PMN-PT.
Mo$^{4+}$ (0.65~\AA) and Tc$^{4+}$ (0.645~\AA) are only slightly larger than Ti$^{4+}$, and they are very weakly FE-active elements, as will be explained later.
Therefore, they exhibit very small displacements, and the Mo/Tc-O bond lengths are all very close to each other (Figs.S1 (g) and (h) in SM).
As a result, not only do the Pb displacement amplitudes decrease, but also the scattering between the Pb displacements increases, especially in PMN-PT-Tc.
This indicates that the introduction of Mo and Tc into PMN-PT increases the disorder of the systems.
The average Pb displacement direction is still around [323] in PMN-PT-Mo, while it changes slightly to [322] in PMN-PT-Tc.

\begin{figure}[htb!]
 \centering
 \includegraphics[width=0.95\columnwidth]{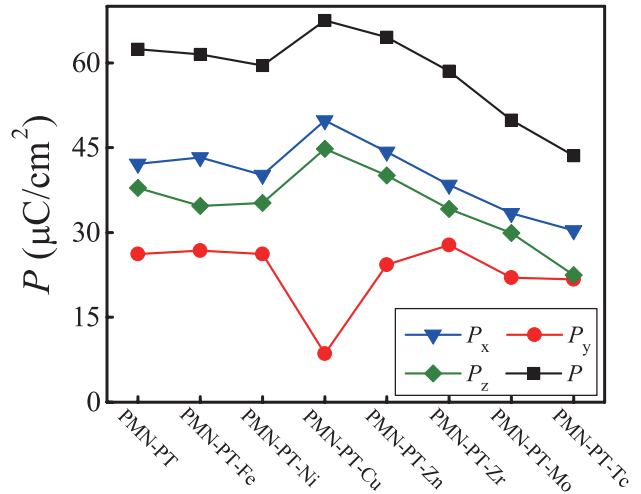}
 \caption{\label{Fig2}(Color online) The amplitudes of the polarization ($P$) and its components ($P_x$, $P_y$, $P_z$) of the eight systems are shown.}
\end{figure}

We have also calculated the polarization of all systems, as shown in Table I.
The polarization and the components are shown in Fig.2.
It is clear that the polarization magnitudes show small deviations from that of PMN-PT, except for the Cu-, Mo- and Tc-doped PMN-PT, for which the polarization magnitudes are notably different.
This behavior is consistent with the similarity and discrepancy of the displacements of the cations in these systems.
As explained above, Fe and Ni have a very limited effect on the local structure, and the polarizations of these two doped systems also show very small changes relative to the value for PMN-PT.
For PMN-PT-Cu, the amplitude of the polarization is increased by 8\% relative to the 62.4 $\mu$C/cm$^2$ polarization of PMN-PT.
It is well-known that the Pb lone-pair 6$s$ electrons give rise to a large Pb displacement and a substantial Pb contribution to the overall polarization.
The average displacement of Pb in PMN-PT-Cu is indeed larger than that in PMN-PT.
However, the much stronger scatter between Pb displacements in PMN-PT-Cu makes the dipole moments of Pb more disordered.
Furthermore, as shown in Table I, the weighted average displacement ($D_\textrm{FE,B}$) of the FE-active B-site cations in PMN-PT-Cu is very close to that in Fe- and Ni-doped and undoped PMN-PT.
The interplay between these factors leads to the increase of the polarization.

In PMN-PT-Zn, Zn also shows a large displacement and forms covalent bonds with O, but Zn does not significantly affect the Pb displacement directions.
Furthermore, as shown in Table I, the average Pb displacement and weighted average displacement of the FE-active B-site cations in PMN-PT-Zn are very close to those in Fe- and Ni-doped and undoped PMN-PT.
Thus, compared to PMN-PT, both the amplitude and the direction of the polarization in PMN-PT-Zn do not show significant changes.
In Zr-doped PMN-PT, the relatively small displacement of the moderately FE-active Zr slightly decreases the weighted average displacement of the FE-active B-site cations.
Due to the combination of the slightly increased Pb displacement and the almost unchanged scatter, the polarization of this system remains almost unchanged.
The situation is quite different in PMN-PT-Mo/Tc.
The polarization magnitudes in PMN-PT-Mo/Tc are greatly decreased.
Due to the less FE-active Mo and Tc ions, $D_\textrm{Pb}$ and $D_\textrm{FE,B}$ are decreased, and the scatter angles between the Pb displacements increase.

\subsection{\label{sec:level2}Electronic structure and charge transfer of three relaxor systems}

\begin{figure*}[htb!]
 \centering
 \includegraphics[width=17cm]{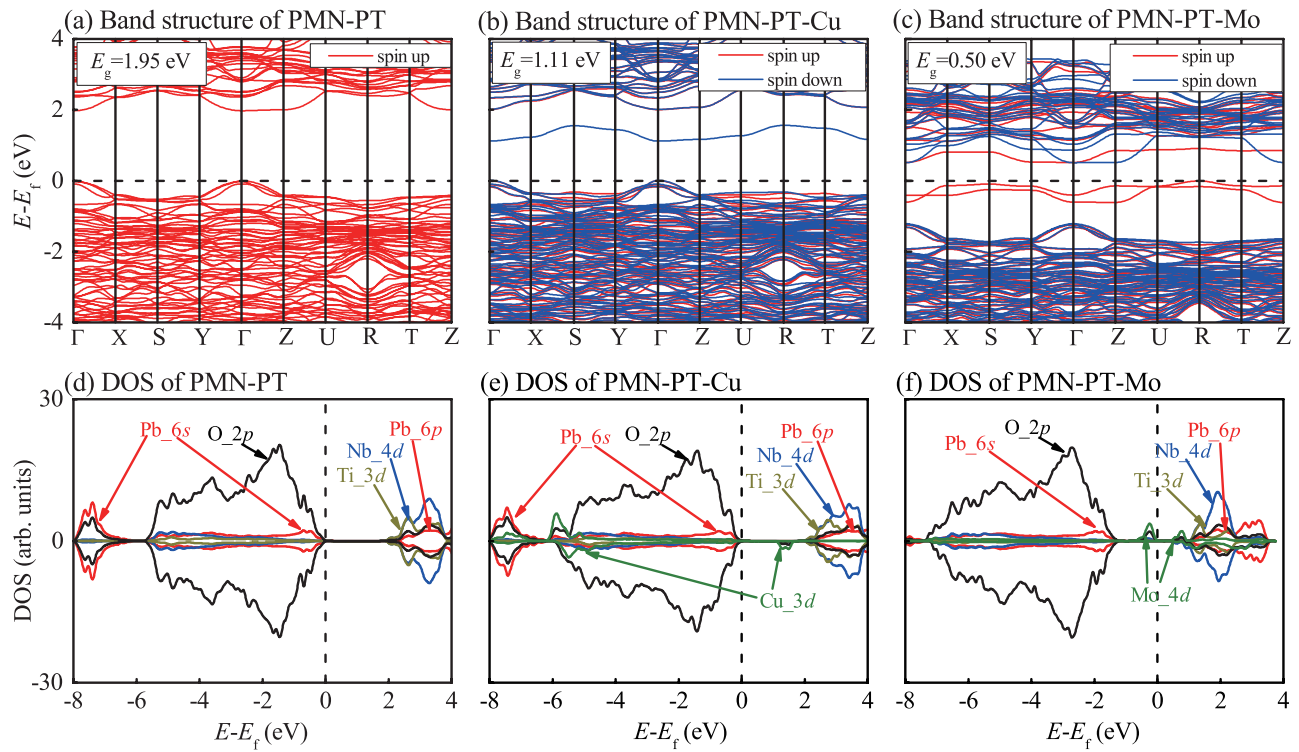}
 \caption{\label{Fig3}(Color online) Electronic structures of three systems: (a) and (d) for PMN-PT, (b) and (e) for PMN-PT-Cu, (c) and (f) for PMN-PT-Mo. The band gap $E_{\textrm{g}}$ values are 1.95, 1.11 and 0.50 eV, respectively. The $E_{\textrm{f}}$ indicates the VBM. For PMN-PT, the valence bands are mainly composed of O 2$p$ and Pb 6$s$ orbitals, while the conduction bands are mainly composed of Ti 3$d$, Nb 4$d$ and Pb 6$p$. Compared to PMN-PT, Cu 3$d$ lowers the CBM while the Mo 4$d$ raises the VBM, thus decreasing the band gaps of both systems.}
\end{figure*}

To obtain a better understanding of the properties of the system, we carried out electronic structure calculations and Bader charge analyses for three systems: PMN-PT, PMN-PT-Cu and PMN-PT-Mo.
The electronic structures are shown in Fig.3.
The undoped PMN-PT has a direct DFT-LDA band gap of 1.95 eV, higher than the DFT-LDA calculated band gap for PbTiO$_3$ (1.50 eV).
The DFT-LDA calculated band gap of PMN-PT is underestimated compared to the experimental value (3.24 eV, see Ref.\cite{JAP96p1387}), which is a well-known problem of the LDA exchange-correlation functional.
Fig.3 (d) shows that for PMN-PT, the valence bands are mainly composed of the O 2$p$ and Pb 6$s$ orbitals, while the conduction bands are mainly composed of the hybridized Ti 3$d$, Nb 4$d$ and Pb 6$p$.
However, when a dopant atom is introduced, some other orbital characters will be introduced into the electronic structure.
For example, when Cu is doped into PMN-PT, as shown in Figs.3 (b) and (e), a Cu 3$d$ band appears in the band gap region of PMN-PT.
This drops the position of the conduction band minimum (CBM), and thus the band gap decreases to 1.11 eV.

Doping of Mo has a very different effect on the electronic structure.
It can be seen from Fig.3 (f) that the valence band maximum (VBM) position is pulled up by the Mo 4$d$ states that contribute to both the valence and conduction bands.
As a result, the band gap is reduced to 0.50 eV.
A further examination of the projected density of states (DOS) of this system shows that the bands immediately below the VBM are mainly composed of $d_{xy}$ and $d_{yz}$ of Mo.
Although the DOS in Fig.3 (f) exhibit hybridizations between $t_{2g}$ $d$ electrons of Mo and $p$ of O, the attractive interaction of Mo to its neighboring oxygen becomes weak, and the octahedron distortion also becomes small.
Consequently, the displacement of Mo is greatly reduced compared to Ti, and the Mo-O bond lengths become more uniform, as shown in Fig.S1 (g) of the SM.
This indicates that Mo$^{4+}$ is weakly FE-active and explains why substituting Ti with Mo decreases the displacements greatly, as described above.

\begin{table}
\caption{Charge transfer of three relaxor systems. X stands for the dopant ion in each doped PMN-PT. The charge unit is electron ($e$).}
\renewcommand\arraystretch{1.30}
\begin{ruledtabular}
\begin{tabular}{lcccccccccccc}
        & Pb & Mg & Nb & Ti & O & X \\
\hline
PMN-PT   & 1.34 & 1.73 & 2.56 & 2.11 & -1.19 & - \\
PMN-PT-Cu & 1.34 & 1.73 & 2.53 & 2.11 & -1.17 & 1.19 \\
PMN-PT-Mo & 1.34 & 1.73 & 2.56 & 2.11 & -1.20 & 2.34 \\
\end{tabular}
\end{ruledtabular}
\end{table}

We now turn to the Bader charge analyses of the systems.
Table II shows the average charge transfer of all ions in the three systems.
In PMN-PT, Mg has the transfer of 1.73$e$ and the Ti transfer is 2.11$e$.
However, when one Mg is replaced by one Cu, the charge transfer of the dopant Cu is only 1.19$e$, 0.54$e$ smaller than that of Mg.
This supports the idea that Cu is more polarizable than Mg, enabling covalent Cu-O bonding.
By contrast, the situation is different in PMN-PT-Mo.
When one Ti atom is substituted by one Mo, the charge transfer of Mo is 2.43$e$, larger than the 2.11$e$ of Ti.
This is consistent with the weaker attractive interaction of Mo and O.
As a result, the Mo-O bond lengths are more uniform and the displacements of Mo becomes very small (Table I).
These charge analyses, as well as the electronic structure, shed light on why the local structures are so different in the doped systems.

\subsection{\label{sec:level2}The $T_{\textrm{m}}$ of the maximum dielectric constant and its dispersion}

For piezoelectric applications, the temperature ($T_{\textrm{m}}$) at which dielectric constant is maximized is one of the most important factors that must be considered.
According to previous work \cite{Ilya04p220101}, the temperature $T_{\textrm{m}}$ in Pb-based relaxor ferroelectrics is proportional to the square of the polarization.
Using equation (3) in Ref.\cite{Ilya04p220101}, we estimated the $T_{\textrm{m}}$ values at 1 MHz for all systems, with the results shown in Table I \cite{note1}.
We can see that even small changes of polarization will give rise to relatively large differences in $T_{\textrm{m}}$ values. For example, the $T_{\textrm{m}}$ of the Zn-doped PMN-PT is dozens of kelvins higher than that of undoped PMN-PT, while the polarization of PMN-PT-Zn is only increased by 2.1 $\mu$C/cm$^2$.
PMN-PT-Cu has the highest $T_{\textrm{m}}$ that is about 80 K higher than the $T_{\textrm{m}}$ (463 K) of PMN-PT, representing a significant $T_{\textrm{m}}$ enhancement.
Such enhancements in $T_{\textrm{m}}$ make PMN-PT-Cu and PMN-PT-Zn promising candidates for applications at higher temperature.
The other dopants either have limited effects on $T_{\textrm{m}}$ or decrease the temperature.
For example, our calculations show that Mo is a very weakly FE-active element and its displacement is much smaller than that of Ti.
The consequently reduced polarization leads to the reduced $T_{\textrm{m}}$ of this doped system, making it a poor candidate for high-temperature applications.

We now qualitatively discuss the effect of the doping on the temperature dispersion.
Experiments \cite{Colla98p3298, MSEB111p107, JACS84p1281} have shown that many lead-based solid solutions exhibit increased temperature $T_{\textrm{m}}$ with higher PT content, while the temperature dispersion $\Delta$$T_{\textrm{m}}$ becomes smaller.
This behavior indicates a certain relationship between $\Delta$$T_{\textrm{m}}$ and $T_{\textrm{m}}$, i.e. the larger the $T_{\textrm{m}}$ values are, the smaller the $\Delta$$T_{\textrm{m}}$ will be (and vice versa).
Our results show that the larger the displacement, the larger the polarization, and the higher the temperature $T_{\textrm{m}}$ of the maximum dielectric constant as well.
Our previous work \cite{Ilya07p267603} showed that the temperature dispersion decreases with the increasing of the average B-site cation displacement.
Thus, in our doped systems, we predict that inclusion of Cu and Zn into PMN-PT will decrease the temperature dispersion $\Delta$$T_{\textrm{m}}$ and weaken the relaxor features, while inclusion of Mo and Tc into PMN-PT will increase the dispersion and lead to much stronger relaxor behavior.

\section{\label{sec:level1}CONCLUSION}

In our studies, we have considered several different substitutions.
The magnetic elements Fe and Ni show limited effect on the local structure and polarization, and the temperature $T_{\textrm{m}}$ is also not increased.
However, magnetic doping is always interesting because such typical magnetic dopants may confer other interesting properties to the system.
For example, in recent years, the strong magnetoelectric coupling effect in the solid solution of BiFeO$_3$ and PZT has been a focus of studies \cite{APL08P192915, NanoscRL12p7:54, JAP14p224106}.
The inclusion of the typical anti-ferroelectric PbZrO$_3$ into PMN-PT also does not lead to enhancement of the $T_{\textrm{m}}$.
However, as one of the mostly studied structures, PbZrO$_3$ shows many anomalous properties under different conditions such as hydrostatic pressure \cite{PRB89p214111}.
Therefore, it would be expected that PbZrO$_3$ could lead to other property changes upon addition to PMN-PT, especially under hydrostatic pressure or even strain.
The doping of FE-active Cu and Zn into PMN-PT will increase the polarization, especially for Cu doping.
The dopants themselves also exhibit large displacements that are comparable to the Nb displacements.
Such doped PMN-PT may be promising candidates for applications at higher temperature.
The temperature dispersion $\Delta$$T_{\textrm{m}}$ values of these systems decrease due to the small increase of the average B-site cation displacements.
Additionally, Cu doping also makes the system slightly more disordered due to the larger scatter angle $\theta$$_{\textrm{Pb}}$ between the Pb displacements.
We have also performed similar calculations for Cu doped 0.75PZN-0.25PT, and find that Cu doping has a limited effect on this system.
This stems partly from the fact that Zn is already strongly FE-active ion, which gives rise to ordered systems, and substituting Zn with another similar FE-active element (Cu) is expected to show limited effects on the local structure.
This is rather different from the substitution of the very weakly FE-active Mg with an FE-active Cu/Zn in PMN-PT.
Thus, we predict that Cu doping can increase the $T_{\textrm{m}}$ only in PMN-PT.
Mo- and Tc-doped PMN-PT show greatly reduced cation displacements and thus reduced polarization values.
As a result, the temperatures $T_{\textrm{m}}$ of the maximum dielectric constant of these systems are much lower than that of PMN-PT.
However, in contrast to the trend for $T_{\textrm{m}}$, the $\Delta$$T_{\textrm{m}}$ of these two doped systems will be increased, implying that these solid solutions are stronger relaxors.
These predictions regarding the relaxor behaviors provide guidance for experimental materials design, especially for the complex materials that have been completely unexplored.

\section{\label{sec:level1}Acknowledgments}

H.X.T. would like to thank Dr. Yuanchang Li for stimulating discussions.
H.X.T. would also like to thank Tsinghua University for the Scholarship for Overseas Graduate Studies.
H.X.T., C.X. and W.D. acknowledge the support from the National Natural Science Foundation of China (Grant No. 11674188) and the Ministry of Science and Technology of China (Grant No. 2017YFB0701502).
H.T. acknowledges the support of the US Department of Energy, under grant DE-FG02-07ER46431.
A.M.R. acknowledges the support of the Office of Naval Research, under grant  N00014-17-1-2574.
The authors acknowledge computational support from the High-Performance Computing Modernization Office of the Department of Defense and the National Energy Research Scientific Computing Center of the Department of Energy.


\begin{thebibliography}{50}
\bibitem{note1} For undoped PMN-PT, the experimental $T_{\textrm{m}}$ value is 397 K (see Ref.\cite{Ilya04p220101} and reference therein). Here our value is 463 K, 66 K higher than the experimental value. Then by interpolating and extrapolating the data of Lei $et$ $al$. \cite{MSEB111p107} for PMN-PT-Ni (i.e. PMN-PNN-PT therein), we estimate the $T_{\textrm{m}}$ to be about 350 K, which is about 70 K lower than our value for PMN-PT-Ni (421 K). So it may be a reasonable estimation that our calculated values are about 70 K higher than the corresponding experimental values.
\end{thebibliography}
\end{document}